\documentclass[reprint,aps,prb,amsmath,amssymb,twocolumn,showkeys, superscriptaddress,floatfix,showkeys]{revtex4-2}
\usepackage{graphicx}
\usepackage{dcolumn}
\usepackage{bm}
\usepackage{bbm} 
\usepackage[colorlinks = true, allcolors = blue]{hyperref}
\usepackage{verbatim}
\usepackage{soul}
\usepackage{float}
\usepackage{xcolor}
\usepackage{multirow}
\usepackage{orcidlink}
\usepackage{chemformula} 

\begin{document}

\title{Twisting in  h-BN bilayers and their angle-dependent properties}

\author{Diem Thi-Xuan Dang\,\orcidlink{0000-0001-7136-4125}}
\email{dangt1@usf.edu}
\thanks{Corresponding authors.}
\affiliation{Department of Physics, University of South Florida, Tampa, Florida 33620, USA}

\author{Dai-Nam Le\,\orcidlink{0000-0003-0756-8742}}
\email{dainamle@usf.edu}
\homepage{https://sites.google.com/view/dai-nam-le/}
\affiliation{Department of Physics, University of South Florida, Tampa, Florida 33620, USA}

\author{Lilia M. Woods\,\orcidlink{0000-0002-9872-1847}} 
\email{lmwoods@usf.edu}
\thanks{Corresponding authors.}
\homepage{https://www.amd-woods-group.com/}
\affiliation{Department of Physics, University of South Florida, Tampa, Florida 33620, USA}

\date{\today}

\begin{abstract}
In this paper, we systematically investigate the structural and electronic properties of twisted h-BN bilayers to understand the role of the twisting angle. Using first-principles methods with relaxation taken into account, we simulate h-BN bilayers with commensurate supercells with the smallest angle being $2.88^{\circ}$ until $60^{\circ}$. We find that the interlayer separation is not constant throughout each bilayer because of the various stacking patterns of AA, AA', AB, AB', and A'B throughout the layers, which also play a significant role in their unique charge redistribution. The calculations for the 110 generated structures show the existence of flat bands in several twisted h-BN bilayers, as well as the emergence of different trends in their properties as a function of the twist angle. 
These results are useful for establishing a systematic base line of registry-dependent relations for the development of more advanced computational methods to access incommensurate h-BN bilayers.

\end{abstract}


\maketitle

\section{\label{sec:1} Introduction}
Layered materials formed by stacking monolayered constituents have been established as a fruitful playground for new properties and technological applications \cite{Novoselov2016}. Due to weak interlayer interactions, the formation of low-dimensional heterostructures has emerged as a way to create tunable materials with fundamental properties different from their constituents \cite{Liu2016}. The registry dependence of different stacking patterns has also been of great interest recently, especially for Moiré materials, in which different patterns induce unconventional property tunability \cite{Nuckolls2024}. For example, flat bands appear in ``magic angle'' twisted bialyer graphene, which can give rise to superconducting, topological, and nematic states due to much enhanced correlation effects \cite{Bistritzer2011,Balents2020}. In twisted semiconducting monolayers, such as transition-metal dichalcogenides, Moiré patterns affect their exciton lifetimes, the emitted light and other optical properties \cite{Choi2021,Tran2019}.

Hexagonal Boron Nitride (h-BN) and its layered variations have also received a great deal of scientific interest. Due to its large energy gap ($>6$ eV), bulk h-BN can be used as a gating element in electronics applications, as a source of single-photon emission and UV radiation, among others \cite{Britnell2012,Sponza2019,Tran2016}. Experiments have shown that Moiré h-BN systems can exhibit tunable photoluminescence spectra and energy gaps \cite{Zhao2020}.

Atomistic simulations for Moiré h-BN bilayers for small twist angles are challenging because of the large number of atoms in the supercells. Most computational studies have focused on a limited selection of twist angles to elucidate enhanced excitonic properties and flat-band formation. For example, using density functional theory (DFT) interfaced with tight-binding models, flat defect-like bands have been shown to occur in h-BN bilayers for sufficiently small twist angles \cite{Zhao2020}. Other studies using similar methods have also shown that flat bands can exist in twisted h-BN bilayers without the need for magic angles \cite{Xian2019,Sponza2024}. First-principles studies combining DFT, effective Hamiltonian models, and quasiparticle band structure calculations have also been employed to demonstrate enhanced excitonic effects on optical spectra of Moiré h-BN systems with a select range of relatively small twist angles \cite{Roman-Taboada2023,Paleari2018,Ninhos2024}.

However, despite this progress, a comprehensive examination of a wide range of commensurate stacking patterns in twisted h-BN bilayers within the capabilities of DFT is not available. Establishing a base-line registry-dependent structure-property relations is crucial for more advanced methods that rely on DFT to capture electron correlations and excitonic effects in h-BN systems. Tracking the role of a wide range of twist angles provides information on preferred registry directions and energy barriers of different metastable states. The effect of the interlayer distance is also an important one to be resolved completely by {\it ab initio} methods. In Moiré systems, the separation between the layers is not constant; thus, full relaxation is needed to capture this situation. The variable spacing between layers as a function of the different stacking patterns is important to determine the characteristic energies and the properties of the electronic structure in Moiré materials \cite{Zhou2015,Li2024}. Because h-BN is a piezoelectric material, it is expected that the lattice relaxation is quite relevant for the distribution of atomistic charges in the bilayered system.

In this paper, we utilize DFT methods to systematically simulate commensurate twisted bilayers by taking into account full structural relaxation and van der Waals (vdW) interactions. The constructed bilayers encompass a relatively large number of systems that are all accessible via DFT methods. Here, the structural and electronic properties are obtained and analyzed in order to understand the role of the twist angle in a larger pool of twisted h-BN bilayers. 

\section{\label{sec:2} Methods}
In this study, DFT calculations are performed using the Vienna Ab initio Simulation Package (VASP) \cite{Kresse1996}. We use a slab structure to model the 2D nature of a twisted bilayer system. Each slab is separated by vacuum regions of 18 \AA \text{ } to avoid impacts from neighboring periodic images. The Perdew-Burke-Ernzerhof exchange correlation functional is used for all calculations \cite{Perdew1996} and vdW interactions are considered within the DFT-D3(BJ) method \cite{Grimme2011}. The plane-wave energy cutoff is 450 eV. All atoms are relaxed until the energy difference of successive atom configurations is less than $10^{-5}$ eV. Geometry optimization is carried out until the residual force in each atom is less than 0.01 eV/\AA. The post-processing of the calculated data is carried out using the VASPKIT code \cite{Wang2021}. For effective mass calculations, the number of points used for the fitting of quadratic functions is 6 with the k-cutoff of 0.001 \AA$^{-1}$. The lattice visualizations are obtained using the VESTA program \cite{Momma2011}.

\section{\label{sec:3} Lattice Structure and registry dependence}

To initiate the simulations, we consider the supercell construction of the twisted bilayers for various twist angles. Monolayer h-BN ($D_{3h}$ symmetry) has a structure very similar to that of graphene; however, their most stable bulk configurations differ: graphite crystallizes in AB stacking order, while bulk h-BN prefers AA' stacking \cite{PEASE1950,Constantinescu2013}. Taking into account the hexagonal symmetry and the two types of atoms that make up the h-BN monolayer unit cell, different initial bilayers are constructed. Starting with the most stable AA' configuration, the bilayer unit cell has BN hexagons positioned above each other with B over N and N over B atoms. The AA configuration is similar to AA'; however, the hexagons lie above each other, with B above B and N above N atoms. By rotating the top h-BN layer about the bottom h-BN layer, the twisted h-BN bilayers are constructed. The rotations are initiated from an AA' and an AA starting registry.

The unit cell of the AA' h-BN bilayer can be described with lattice vectors $\mathbf{a}_1=a(1,0)$ and $\mathbf{a}_2=\frac{a}{2}(\sqrt{3},1)$ in the two layers with $a = 2.50$ \AA \text{} being the lattice constant. The superlattice of the commensurate twisted bilayers can be obtained following similar rules for the twisted bilayer graphene nomenclature \cite{Moon2013,Aragon2019}. The twist is described by translation vectors $L_1 = n\mathbf{a}_1 + m\mathbf{a}_2$ and $L_2 = -n\mathbf{a}_1 + (n + m)\mathbf{a}_2$, where $(n,m)$ are integers. The rotational angle between the two h-BN monolayers can be expressed as $\cos\theta=\frac{n^2+4nm+m^2}{2(n^2+nm+m^2)}$. Starting with AA' stacking ($\theta=0$) and rotating about the center of the BN hexagon in this unit cell, we generate 37 commensurate twisted h-BN bilayers. The smallest angle $\theta=2.88^{\circ}$ with $ n= 11,m = 12$ has $N_{at} = 1588$, the largest number of atoms in the supercell among all combinations. At $\theta=60^{\circ}$, one recovers the AA h-BN stacking. Similarly, starting from the AA bilayer and rotating about the center of the hexagon, 35 commensurate supercells are generated upon rotation by $\theta$. At $\theta=60^{\circ}$, we recover the AA' h-BN bilayer. Additional 36 systems are also generated by starting from the AA' bilayer unit cell and rotating about the B/N atom in the cell. In this case, $\theta=60^o$ results in a bilayer with similar to graphene AB stacking. In addition to AB' and A'B, we simulate 110 twisted h-BN bilayers in total.

\begin{widetext}\begin{figure}[H]
    \begin{center}
    \includegraphics[width = 1 \textwidth]{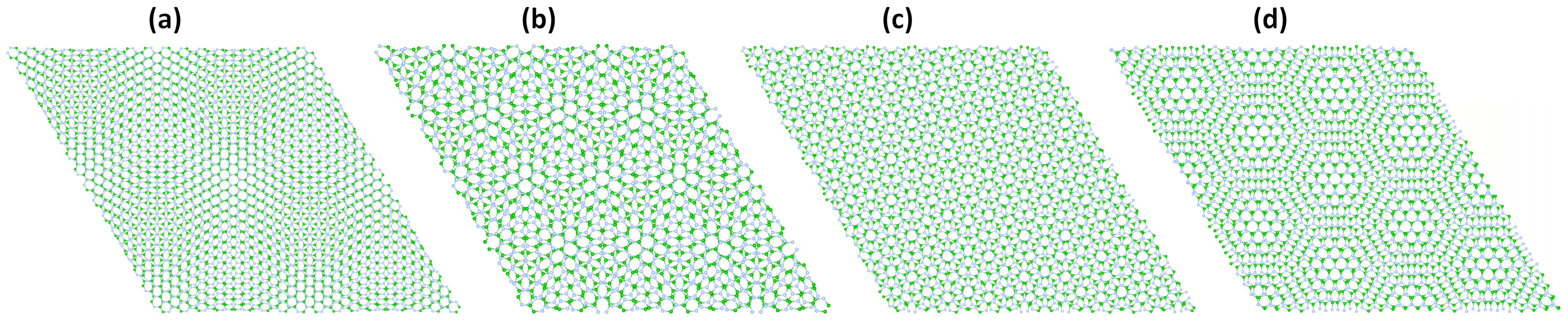}
    \caption{\label{fig:1} Lattice structures of twisted h-BN bilayers obtained from an initial AA' stacking rotating around the center of the BN hexagon with twist angles: (a) $3.89^{\circ}$; (b) $10.90^{\circ}$, (c) $30.16^{\circ}$, (d) $52.07^{\circ}$. }
    \end{center}
\end{figure}\end{widetext}

Several twisted bilayers are shown in Fig. \ref{fig:1} for rotations around the center of the BN hexagon starting from AA'. We find that for $0<\theta<20^{\circ}$, there are distinct patterns with regions displaying AA', A'B, and AB' registry. Here, we follow the established notation for the AA and AB variations of h-BN stacking as described in Ref. \citenum{Constantinescu2013}. For configurations with $0<\theta<20^{\circ}$ obtained from AA, we distinguish AA and AB patterns. For $\theta\sim 30^{\circ}$, such patterns are difficult to distinguish regardless of the starting stacking.

\onecolumngrid

\twocolumngrid

The structural stability of each twisted h-BN bilayer is characterized by examination of the interlayer separation. Because of the polarity of this material, lattice relaxation must be taken into account when considering interlayer separation. In a polar material, such as h-BN, the strain causes charge accumulation in different regions, which is closely related to the atomistic registry of the bilayer and its relaxation during simulations.

\begin{widetext}\begin{figure}[H]
    \begin{center}
    \includegraphics[width = 1.0 \textwidth]{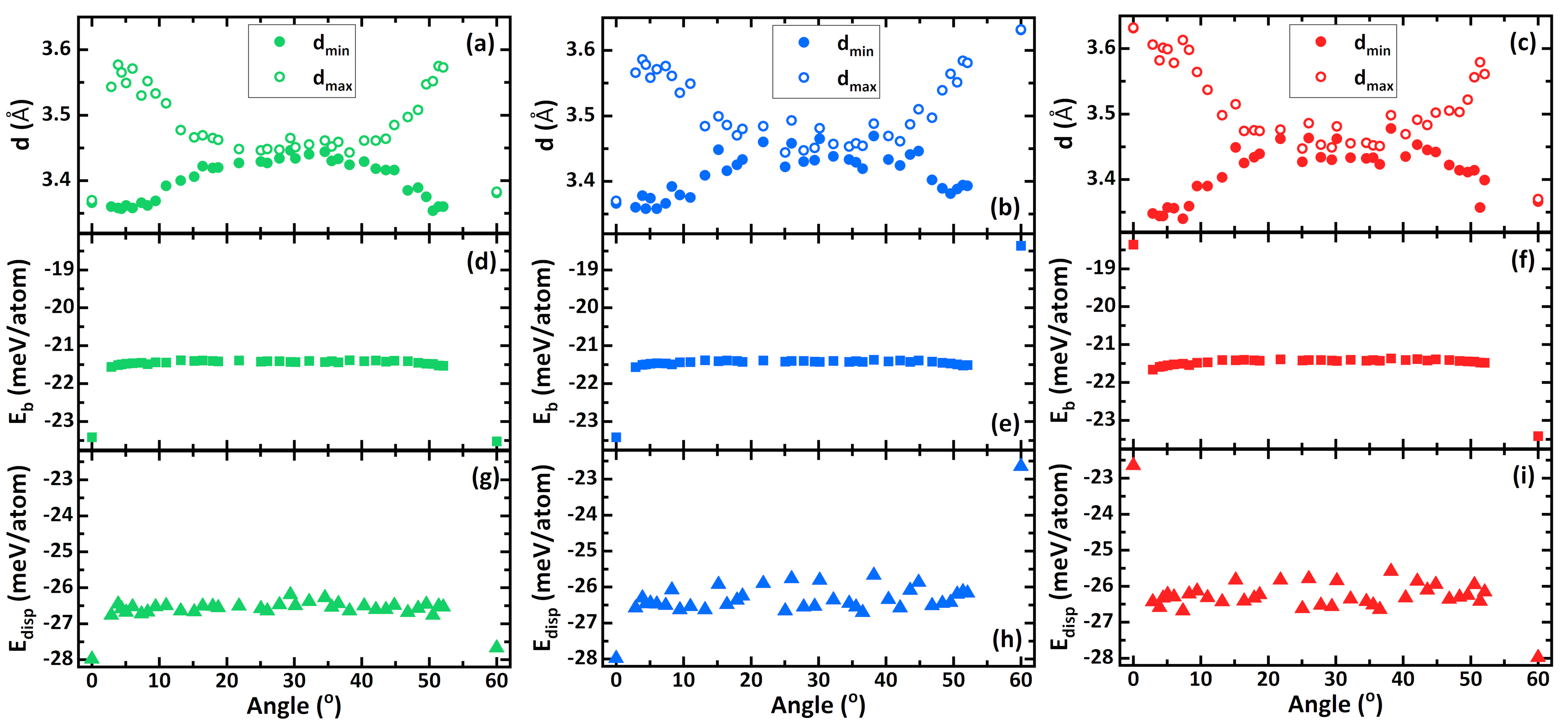}
    \caption{\label{fig:2} The largest $d_{max}$ and smallest $d_{min}$ interlayer separations in twisted h-BN bilayers as a function of the twist angle $\theta$ for: AA' initiated stacking patterns rotating around (a) the B/N atom and (b) the center of the hexagon; and (c) AA initiated stacking patterns rotating around the center of the hexagon. The interlayer binding energy $E_b$ as a function of $\theta$ for AA' initiated stacking patterns rotating around (d) the B/N atom and (e) the center of the hexagon; and (f) AA initiated stacking patterns rotating around the center of the hexagon. The interlayer dispersion energy $E_{disp}$ as a function of $\theta$ for AA' initiated stacking patterns rotating around (g) the B/N atom and (h) the center of the hexagon; and (i) AA initiated stacking patterns rotating around the center of the hexagon.}
    \end{center}
\end{figure}\end{widetext}

We find that the interlayer separation is not constant throughout each configuration. In Fig. \ref{fig:2}(a,b,c), we show the largest $d_{max}$ and the smallest $d_{min}$ separations in each structure. The largest difference between $d_{max}$ and $d_{min}$ is found for twist angles with a large number of atoms in the supercell displaying Moiré patterns with large lattice vectors. The bilayer separation in the AA (or A'B and AB') regions in these Moiré patterns corresponds to $d_{max}$, while the bilayer separation in the AA' (or AB) regions corresponds to $d_{min}$. For example, for $\theta=2.88^{\circ}$, we obtain $d_{max} - d_{min} = 0.258$ \AA \text{ } for the AA stacking and $d_{max} - d_{min} = 0.206 $ \AA \text{ } for the AA' stackings (rotation around the center of BN hexagon). For $\theta\sim (20^{\circ}, 40^{\circ})$, on the other hand, $d_{max}-d_{min}$ does not exceed 0.05 \AA \text{ }, which makes the difference much less noticeable. Further increase in $\theta$ results in an increase in the difference $d_{max}-d_{min}$ to $\theta=60^{\circ}$. At that point, the twisting of AA' around the B/N atom reaches the metastable configuration of AB with $d_{min} = d_{max} = 3.382$ \AA \text{} (Fig. \ref{fig:2}(a)), while the configuration of AA generated by rotation around the center of the hexagon reaches $d_{min} = d_{max} = 3.632$ \AA \text{} (Fig. \ref{fig:2}(b)). The twisting of AA around the center of the hexagon leads to the metastable state of AA' with $d_{min} = d_{max} = 3.368$ \AA \text{} (Fig. \ref{fig:2}(c)).

\onecolumngrid

\twocolumngrid

\begin{widetext}\begin{figure}[H]
    \begin{center}
    \includegraphics[width = 1.0 \textwidth]{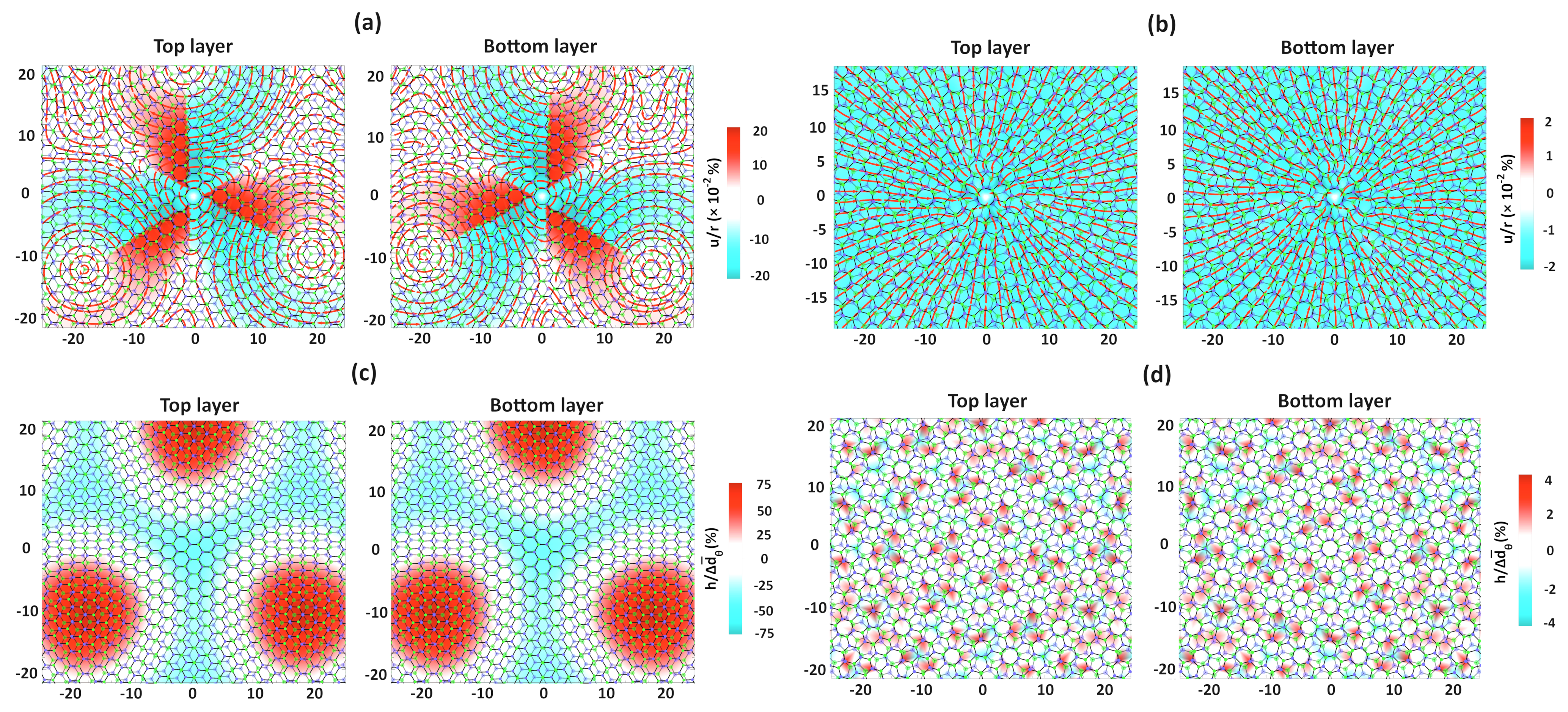}
    \caption{\label{fig:3} Contour maps of the ratio $\mathbf{u}/|\mathbf{r}|$ of twisted h-BN bilayers obtained from an initial AA' stacking rotating around the center of the hexagon with twist angles: (a) $3.89^{\circ}$; (b) $30.16^{\circ}$. The red and cyan colors represent the deformation of the in-plane strain and stress. Contour maps of the ratio $h/\Delta \overline{d}_0$ for the same type of twisted h-BN bilayers with twist angles: (c) $3.89^{\circ}$; (d) $30.16^{\circ}$. The red and cyan colors for the map of the out-of-plane deformation represent the outward (increase interlayer distance) and inward (reduce interlayer distance) types of out-of-plane deformation. }
    \end{center}
\end{figure}\end{widetext}

The variable interlayer distance of the different structures indicates that the h-BN monolayers experience in-plane and out-of-plane deformations. To get a better idea of the in-plane deformation of the individual h-BN monolayers upon twisting, we track the ratio $\mathbf{u}/|\mathbf{r}|$, where $\mathbf{u}$ is the atomic displacement before and after relaxation while $\mathbf{r}$ is the position vector of each atom from the center of rotation. In Fig. \ref{fig:3}(a,b), the contour maps of this ratio are shown for twist angles $3.89^{\circ}$ and $30.16^{\circ}$ of bilayers obtained from rotations about the center of the BN hexagon.  The red and cyan colors represent strain ($\mathbf{u} \cdot \mathbf{r} > 0$) and stress ($\mathbf{u} \cdot \mathbf{r} < 0$) types of in-plane deformation. For $\theta = 3.89^{\circ}$, the top and bottom layers have alternating regions of strain and stress. For $\theta = 30.16^{\circ}$, both layers shrink but in different directions. 

In Fig. \ref{fig:3}(c,d), we also show the contour maps of the ratio $h/\Delta \overline{d}_0$, in which $h$ is the local height of each layer (along the $c$ axis) while $\Delta \overline{d}_0$ is the change in the average interlayer distance before and after relaxation. Tracking this ratio shows the out-of-plane deformation that each layer experiences upon relaxation. The red and cyan colors represent outward (increased interlayer distance) and inward (reduced interlayer distance) types of out-of-plane deformation upon relaxation. The out-of-plane deformations of the twisted h-BN bilayers at $\theta = 3.89^{\circ}$ for the top and bottom layers are similar, indicating that the two layers experience similar corrugation, as also found by Ref. \citenum{Uchida2014} for twisted bilayer graphene. The out-of-plane deformation of the twisted h-BN bilayers at $\theta = 30.16^{\circ}$ for the top and bottom layer, however, show that the amplitude of the corrugation is much smaller, making it difficult to discern long-ranged patterns.

\onecolumngrid

The structural stability of the twisted h-BN layers is also closely related to their interlayer binding energy calculated as $E_b=E_{tot}-2E_{mono}$ ($E_{tot}$ - total DFT energy for the bilayer; $E_{mono}$ - total DFT energy for the monolayer). Figs. \ref{fig:2}(d,e,f) show that the binding energy for the AA' configuration is $E_b=-23.42$ meV/atom as opposed to the AA configuration with $E_b=-18.37$ meV/atom.  The binding energies of the AB, A'B and AB' configurations are $-23.56$, $-18.88$ and $-22.65$ eV, respectively. It is interesting to note that the binding energy for all other systems with $0^{\circ}<\theta<60^{\circ}$ is rather constant hovering around $E_b\sim-21.4$ meV/atom. Looking at the results of dispersion energy, it is found that the largest dispersion energy is in the AA' configuration $E_{disp} = -27.99$ meV/atom and the weakest in the AA configuration $E_{disp} = -22.65$ meV/atom (Fig. \ref{fig:2}(g,h,i)). Larger spreads in values are found for the rest of the systems with $E_{disp}\sim(-26,-27)$ meV/atom for AA' rotating around the B/N atom and $E_{disp}\sim(-25,-27)$ meV/atom for AA' or AA rotating around the center of the hexagon.

\section{\label{sec:4} Electronic Structure}

\begin{widetext}\begin{figure}[H]
    \begin{center}
    \includegraphics[width = 1 \textwidth]{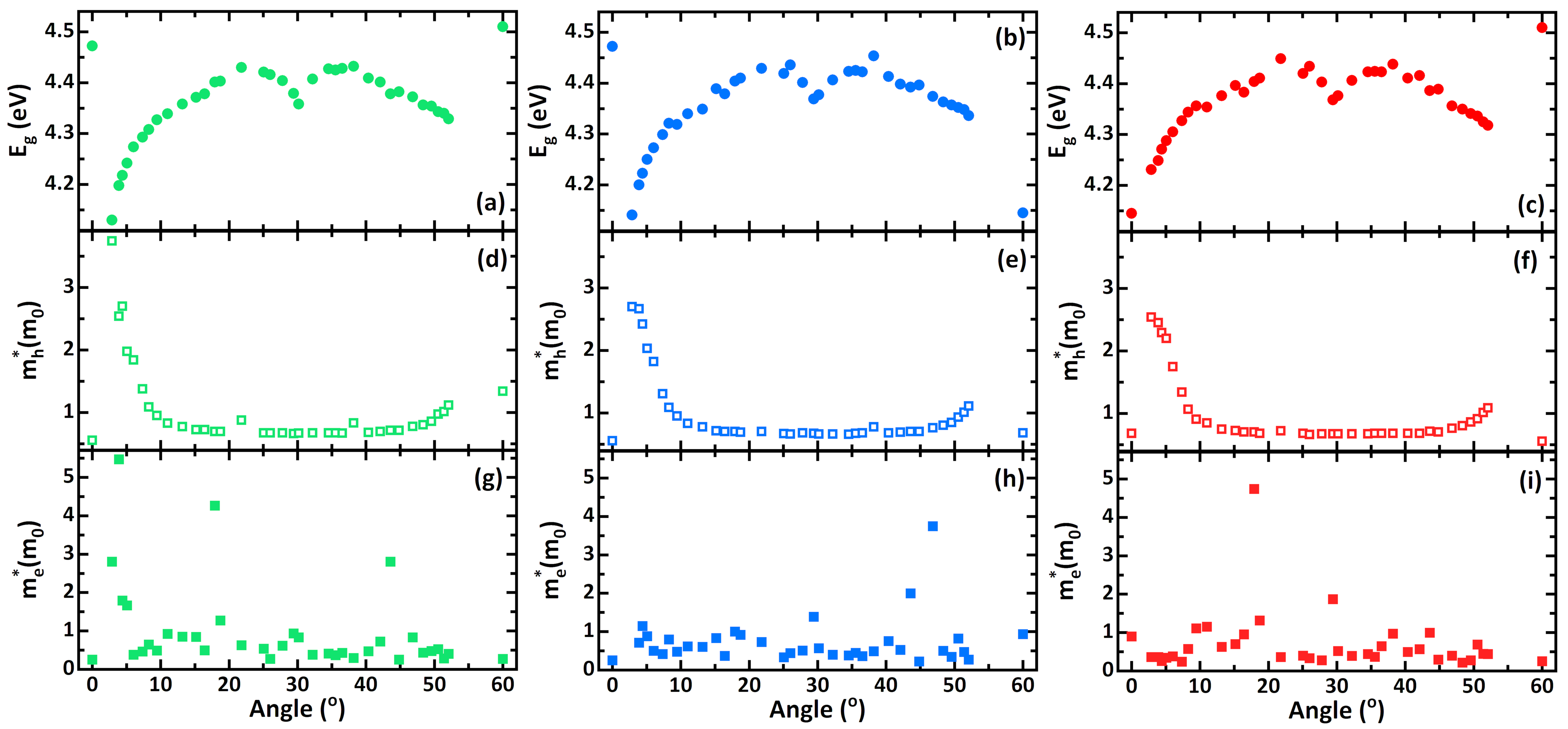}
    \caption{\label{fig:4} Energy band gaps $E_g$ in twisted h-BN bilayers as a function of the twist angle $\theta$ for: AA' initiated stacking patterns rotating around (a) the B/N atom and (b) the center of the hexagon; and (c) AA initiated stacking patterns rotating around the center of the hexagon. The minimum effective hole masses $m^*_h$ as a function of $\theta$ for: AA' initiated stacking patterns rotating around (d) the B/N atom and (e) the center of the hexagon; and (f) AA initiated stacking patterns rotating around the center of the hexagon. The minimum effective electron masses $m^*_e$ as a function of $\theta$ for: AA' initiated stacking patterns rotating around (g) the B/N atom and (h) the center of the hexagon; and (i) AA initiated stacking patterns rotating around the center of the hexagon.}
    \end{center}
\end{figure}\end{widetext}

The effect of the twist angle is further studied in the electronic structure of the h-BN bilayers. All h-BN bilayers are found to be semiconductors, and the evolution of the band gap $E_g$ as a function of the twist angle is shown in Fig. \ref{fig:4}(a,b,c). We find that the most stable AA' stacking has a gap $E_g = 4.472$ eV, which is in reasonable agreement with the previously reported value of 4.650 eV \cite{Henriques2022}. The band gap for the least stable AA configuration is obtained as $E_g=4.145$ eV. Fig. \ref{fig:4}(a,b,c) shows an inverse relation between the twist angle and $E_g$ for $\theta<30^{\circ}$ as increasing the twist angle reduces the band gap. Given the structural symmetry of the AA' and AA faults, the dependence of the band gap as a function of the twist angle also experiences the $\theta \rightarrow 60^{\circ}-\theta $ correspondence - in general, the AA' band gaps for $\theta<30^{\circ}$ will correspond to AA band gaps for $\theta>30^{\circ}$. A slight drop in the trend of the band gap is found around $\theta\sim 30^{\circ}$ as for $\theta=30.16^{\circ}$, the calculated band gap is $E_g = 4.377$ eV for AA' rotating around the center of BN hexagon.

\onecolumngrid

\begin{widetext}\begin{figure}[H]
    \begin{center}
    \includegraphics[width = 1 \textwidth]{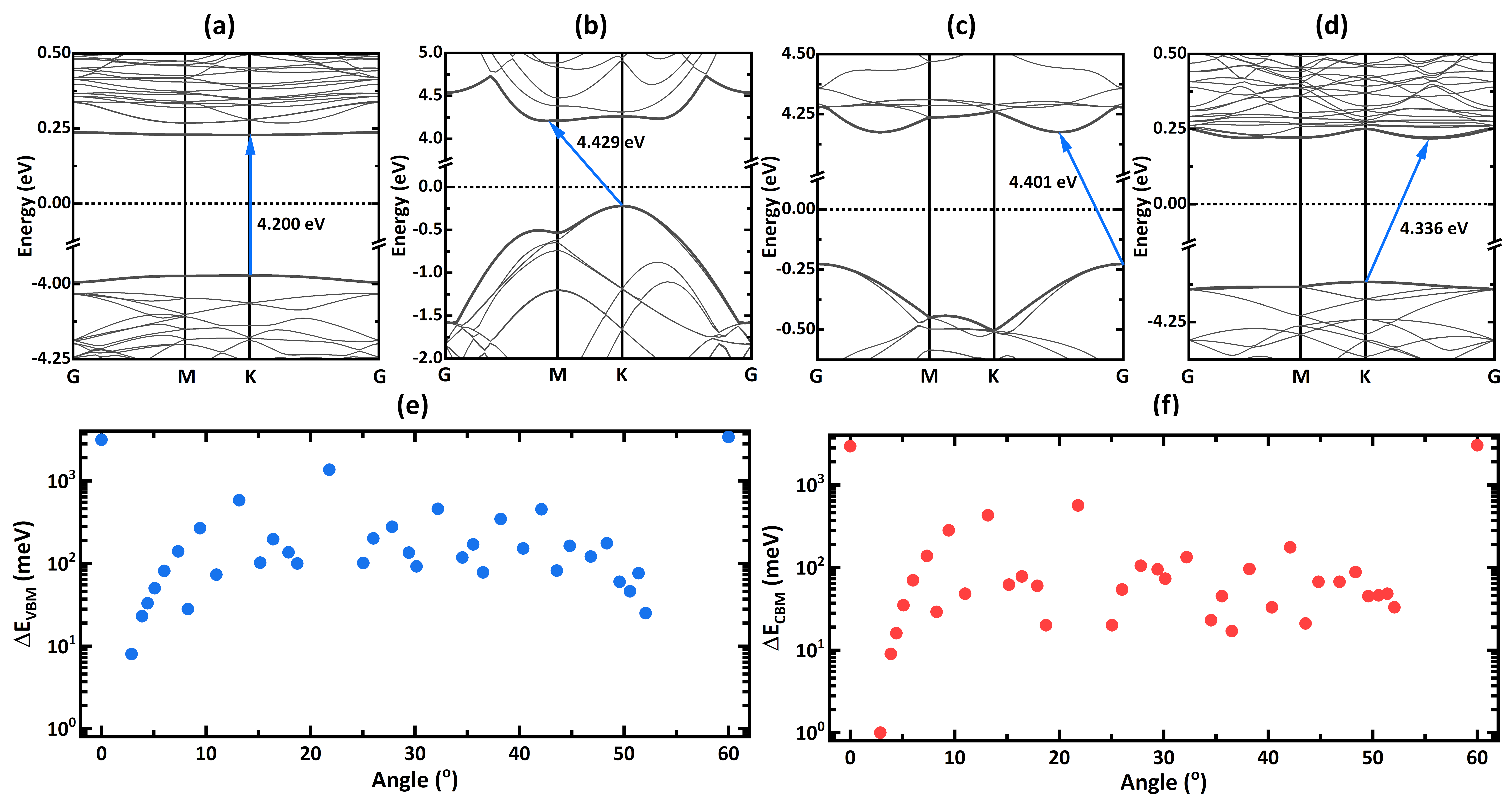}
    \caption{\label{fig:5} The energy band structures of twisted h-BN bilayers obtained from an initial AA' stacking rotating about the center of the BN hexagon with twist angles: (a) $3.89^{\circ}$; (b) $21.79^{\circ}$, (c) $27.80^{\circ}$, (d) $52.07^{\circ}$. The bandwidth of the (e) valence band maximum and (f) conduction band minimum as a function of the twisted angles of the hBN twisted bilayers from AA' stacking rotating around the center of the hexagon.}
    \end{center}
\end{figure}\end{widetext}

In Fig. \ref{fig:5}(a-d), the energy band structures are shown for several twist angles. Flat bands are displayed for $\theta=3.89^{\circ}$ and $\theta=52.07^{\circ}$. In fact, flat bands similar to Fig. \ref{fig:5}(a) are found for all considered twist angles less than $7.34^{\circ}$, while flat bands similar to Fig. \ref{fig:5}(d) are found for angles greater than $44.82^{\circ}$. Apparently, flat bands do not need a special angle as is the case with twisted bilayer graphene, but it is rather an intrinsic property of twisted h-BN bilayers \cite{Zhao2020,Yu2023}. Figs. \ref{fig:4}(a-c) and \ref{fig:5}(a-d) also show that although all systems considered systems are semiconductors, for some $\theta$ the bilayers have direct band gaps, while in others the twisted bilayers appear as indirect semiconductors. 

To track the flatness of the bands, in Figs. \ref{fig:4}(d,e,f), the minimum effective hole masses $m_h^*$ at the maximum point of the highest valence band are shown, and the effective electron masses of the lowest conduction band $m_e^*$ at its minimum point are given in Fig. \ref{fig:4}(g,h,i). The effective mass is calculated using the relation $m^*=\hbar^2 \left(\frac{d^2E(k)}{dk^2}\right)^{-1}$, where $E(k)$ is the dispersion of the corresponding energy band with respect to the wave vector $k$ obtained from the simulations. Figs. \ref{fig:4}(d,e,f) show that for $\theta<10^{\circ}$, the effective hole masses are quite heavy. For example,  $m_h^*=3.743$ $m_0$, $2.701$ $m_0$ and $2.540$ $m_0$ for $\theta=2.88^{\circ}$ for AA' initiated pattern rotating around the B/N atom; AA' and AA initiated patterns rotating around the center of the hexagon, respectively. However, for twist angles in the $(20^{\circ},50^{\circ})$ range, $m_h^*$ is similar to $m_h^*=0.56$ $m_0$ (and $m_h^*=0.68$ $m_0$) for the AA' (and AA) bilayers. The effective electron mass is also relatively heavy, although the most dramatic increases are found for larger twist angles. From Fig. \ref{fig:4}(h), one finds $m_e^*=3.743$ $m_0$ for $\theta = 46.83^{\circ}$ and Fig. \ref{fig:4}(i) gives $m_e^*=4.741$ $m_0$ for $\theta = 17.90^{\circ}$. Comparison of the different panels in Fig. \ref{fig:4} further shows a well-expressed inverse corelation between $E_g$ and $m_h^*$, such that configurations with small band gaps have large effective hole masses. On the other hand, such a clear relation is not found between band gaps and effective electron masses.

The flatness of the highest valence and lowest conduction bands is further examined by calculating their band spreads $\Delta E_{VBM, CBM}$ with the results given in Fig. \ref{fig:5}(e,f). The largest $\Delta E_{VBM} = 3.385$ eV and $\Delta E_{CBM} = 3.015$ eV are obtained for the initial AA' stacking for a twist angle of $\theta=60^{\circ}$\textbf{}. For a twist angle of $\theta=2.88^{\circ}$, $\Delta E_{VBM} = 8$ meV and $\Delta E_{CBM} = 1$ meV for the valence and conduction bands, respectively. Although for $\theta<10^{\circ}$, the band widths are small, similar values are also found for larger angles. For example, the valence band $\Delta E_{VBM} = 0.136$ meV and the conduction band $\Delta E_{CBM} = 0.06$ meV for $\theta = 17.90^{\circ}$. 

\twocolumngrid

\onecolumngrid

\twocolumngrid

\begin{widetext}\begin{figure}[H]
    \begin{center}
    \includegraphics[width = 1 \textwidth]{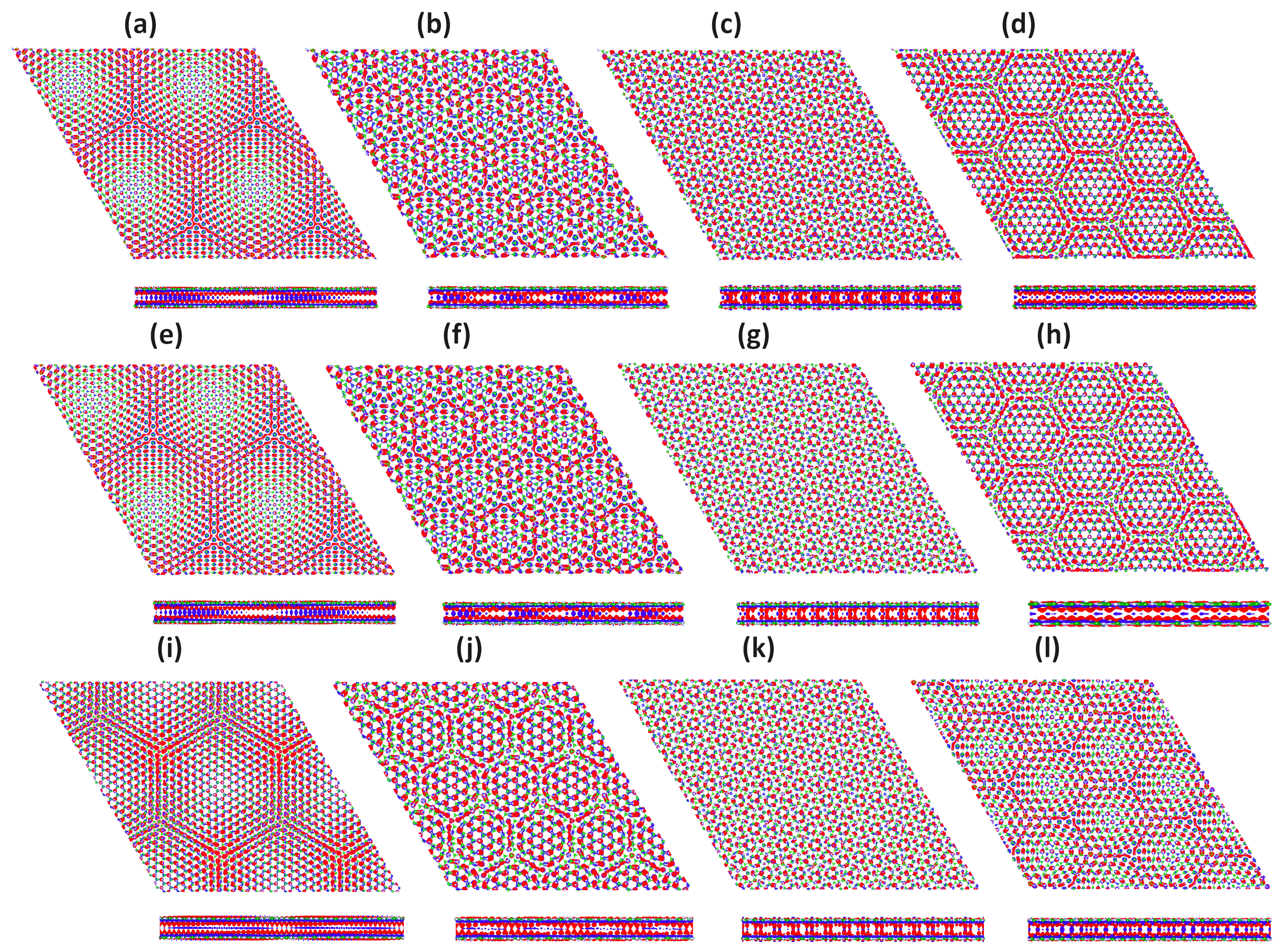}
    \caption{\label{fig:6} Top and side views of the charge density difference in twisted h-BN bilayers obtained from an initial AA' stacking rotating around the B/N atom with twist angles: (a) $3.89^{\circ}$; (b) $10.90^{\circ}$, (c) $30.16^{\circ}$, (d) $52.07^{\circ}$ and the center of hexagon with twist angles: (e) $3.89^{\circ}$; (f) $10.90^{\circ}$, (g) $30.16^{\circ}$, (h) $52.07^{\circ}$, and from an initial AA stacking rotating around the center of hexagon with twist angles: (i) $3.89^{\circ}$; (j) $10.90^{\circ}$, (k) $30.16^{\circ}$, (l) $52.07^{\circ}$.  The blue and red colors denote electron accumulation and depletion, respectively. The isosurface values are set to $7 \times 10^{-5}$ and $10^{-4}$ $e$\AA$^{-1}$ for the top and side views, respectively.}
    \end{center}
\end{figure}\end{widetext}

It is also interesting to consider the evolution of the charge redistribution as the h-BN layers are twisted (see Fig. \ref{fig:6}). The charge redistribution is calculated as the difference in charge density $\Delta \rho = \rho_{total} - \rho_{top} - \rho_{bottom}$, where $\rho_{total}$ represents the total charge density of the twisted bilayer, and $\rho_{top/bottom}$ is the charge density of only one top (bottom) layer.  

Fig. \ref{fig:6} (a, e, i) depict distinct hexagonal patterns for $\theta=3.89^{\circ}$ twisting initiated from AA' and AA stackings: the top view of panel (a, e) show charge depletion in the A'B regions and charge accumulation at and near their boundaries outlined by the AA' and AB' stacking, while panel (i) shows no charge in the AA regions and a zigzag-like sequence of charge depletion and charge accumulation at the boundaries outlined by AB and BA patterns. The side view of the panels shows that the charge redistribution is mainly located in the interfacial region. Distinct regions of charge accumulation and depletion are also found for $\theta=10.90^{\circ}$ and $\theta=52.07^{\circ}$ (Fig. \ref{fig:6}(b,f,j,d,h,l)), however, their spatial extensions are smaller compared to the case of $\theta=3.89^{\circ}$. Also, for $\theta=30.16^{\circ}$, the Moiré patterns are rather smaller, giving very similar charge distributions for the AA' and AA initiated stackings, as shown in Fig. \ref{fig:6}(c,g,k).

\onecolumngrid

\twocolumngrid

\section{\label{sec:5} Conclusions}

In this study, we have considered a relatively large number of commensurate twisted h-BN bilayers, which are simulated using DFT methods by taking into account the full structural relation. Although monolayered h-BN has a hexagonal symmetry as in graphene, its two different atoms result in variations of the graphene-like AA and AB stacking configurations. No matter the initial AA or AA' pattern, twisting by small angles (less that $10^{\circ}$) leads to flat bands in the conduction and valence regions for many registry faults. 

As the AA' or AA initial bilayers are being twisted, we observe clear trends of the evolution of the energy band gap, the effective masses of the valence and conduction bands,  and the band widths. On the other hand, the binding energy is almost a constant for the AA' or AA stacking. Thus, the h-BN bilayer twisting experiences small energy barriers between stackings with $0^{\circ}<\theta<60^{\circ}$ and is rather smooth. The interlayer distance, however, is not constant with changing $\theta$. The larger Moiré patterns have larger variations in $d$, with distinct charge redistribution patterns distinguishing the different twisting angles. We believe that this systematic study gives a good base-line model representation of twisted h-BN bilayers over a wide range of twist angles, which can serve as a stepping stone for other more advanced computations suitable for incomensurate structures.

\begin{acknowledgements}
D.T-X.D. acknowledges support from Presidential Fellowship sponsored by University of South Florida. L.M.W. acknowledges financial support from the US National Science Foundation under grant No. 2306203. Computational resources are provided by USF Research Computing.
\end{acknowledgements}

\section*{Author contribution}
D.T-X.D. performed the simulations and analysis; D.N.L. performed the analysis; L.M.W. conceived the idea, performed the analysis and edited the manuscript.

\bibliography{ref}

\end{document}